\documentclass[paper, 12pt, letterpaper, epsf]{JHEP} 
\input{epsf.tex}
\usepackage{graphics, latexsym} 
\usepackage{epsfig}

\def\Bbb{\bf} 

\def\C{{\Bbb C}} 
\def\R{{\Bbb R}}
\def\Z{{\Bbb Z}} 
\def\H{{\Bbb H}}

 \font\mybb=msbm10 at 12pt
 \def\bb#1{\hbox{\mybb#1}}

\def\S {\bb{S}}
 
\def\CC {B}

\def\id{\protect{{1 \kern-.28em {\rm l}}}}

\newcommand{\be}{\begin{equation}} \newcommand{\ee}{\end{equation}}
\newcommand{\bea}{\begin{eqnarray}} \newcommand{\eea}{\end{eqnarray}}
\newcommand{\beann}{\begin{eqnarray*}} \newcommand{\eeann}{\end{eqnarray*}}
\newcommand{\bfig}{\begin{figure}} \newcommand{\efig}{\end{figure}}
\newcommand{\nn}{\nonumber}
\newcommand{\ba}{\begin{array}}\newcommand{\ea}{\end{array}}

\newcommand{\eg}{{\em e.g., }}

\newcommand{\ra}{\rightarrow}

\newcommand{\halfof}[1]{\frac{#1}{2}}

\newtheorem{Proposition}{Proposition}[section]

\newtheorem{Theorem}{Theorem}[section]
\newtheorem{Lemma}{Lemma}[section]
\newtheorem{Corrolary}{Corrolary}[section]

\newcommand{\bp}{\begin{Proposition}} \newcommand{\ep}{\end{Proposition}} 
\newcommand{\bt}{\begin{Theorem}} \newcommand{\et}{\end{Theorem}} 
\newcommand{\bl}{\begin{Lemma}} \newcommand{\el}{\end{Lemma}} 
\newcommand{\bc}{\begin{Corrolary}} \newcommand{\ec}{\end{Corrolary}} 

\title{Massive Gravitino Propagator in Maximally Symmetric Spaces and Fermions in dS/CFT}

\author{L. Anguelova, P. Langfelder
\\C.~N.~Yang Institute for Theoretical Physics\\
SUNY at Stony Brook, NY, 11794-3840,
U.S.A.\\anguelov, plangfel @insti.physics.sunysb.edu}

\abstract{We extend the method of calculation of propagators in maximally symmetric spaces (Minkowski, dS, AdS and their Euclidean versions) in terms of intrinsic geometric objects to the case of massive spin $3/2$ field. We obtain the propagator for arbitrary space-time dimension and mass in terms of Heun's function, which is a generalization of the hypergeometric function appearing in the case of other spins. As an application of this result we calculate the conformal dimension of the dual operator in the recently proposed dS/CFT correspondence both for spin $3/2$ and for spin $1/2$. We find that, in agreement with the expectation from analytic continuation from AdS, the conformal dimension of the dual operator is {\it always} complex (i.e. it is complex for every space-time dimension and value of the mass parameter). We comment on the implications of this result for fermions in dS/CFT.} 

\preprint{YITP-SB-03-04}

\begin{document}



\vskip .3in

\section{Introduction} 

The recent observational data in favor of the presence of a positive cosmological constant in the universe make it
especially important to understand how to formulate consistent theory of all interactions in de Sitter (dS) space. This
is highly nontrivial: Quantum field theory in dS presents us with a lot of puzzles \cite{W, BFP, GKS, Suss0208}, and
whether and how they could be resolved in the underlying fundamental theory is not at all clear, as it is still not
known how to obtain a stable de Sitter solution in the best-so-far candidate for such a unifying theory -- string
theory.\footnote{Recently there has been a progress in that direction: Fr\'e, Trigiante and Van Proeyen \cite{vPr}
found stable de Sitter vacua in $N=2$ supergravity in $4$ dimesnions. However, their embedding in string theory is
still an open problem. Another exciting recent development is \cite{KKLT}, where metastable dS vacua were found in type
IIB string theory.} A key ingredient in the final picture may be holography, which is believed to be an essential
feature for any consistent theory of quantum gravity \cite{tH, Sus}. One realization of this idea is the AdS/CFT
correspondence \cite{JM, GKP, EW}, which has been studied in a huge number of cases during the last few years (for a
review see \cite{AdSrev}). Another is the recently proposed dS/CFT correspondence \cite{AS}. Although it is hoped that
it may shed light on quantum gravity in de Sitter space, the lack of a clear relation to string theory is hindering an
explicit realization of this proposal, and it is largely modeled on analogy with AdS/CFT (see, for example, the
prescription for computation of scalar field correlation functions in the boundary theory \cite{SV}). There have even
been papers arguing that dS/CFT is merely an analytic continuation of AdS/CFT \cite{KS, CC}. One should not forget,
however, the fundamental differences between physics in dS and AdS. For example, the analytic continuation of the
vacuum state in AdS space does not coincide with any of the vacua of de Sitter space\footnote{As known since \cite{ChT,
BA, EM}, there is an infinite one parameter family of de Sitter invariant vacua in dS. Their possible role in the
cosmology of the early universe has been explored in \cite{UD1, UD2, KKLSS, UD, DDO} and references therein. In
addition the question which one is the most reasonable (physical) vacuum state in dS is still unsettled~\cite{BM,
PSV, EL, UD, CMS, GL}.}. Also, unlike AdS, dS has two boundaries, posing the (as yet unsettled) question whether the
dual theory should be thought of as a single CFT \cite{AS, LMM} or two entangled CFT's \cite{BBM}.

Field theory in de Sitter space was studied extensively in the '80s due to interest sparked by inflationary cosmology.
A technique of calculation of propagators in maximally symmetric spaces was developed in a series of papers \cite{AJ,
AL, AT}. The main idea is the following: One chooses a basis of bitensors which are invariant under the symmetry group
of the space under consideration, and makes an Ansatz for the propagator in terms of these bitensors multiplied by
coefficients that are functions only of the geodesic distance. The coefficient functions are then  determined so that
the propagator satisfies the appropriate field equations and constraints. The original papers considered spins $0$ and
$1$ in arbitrary dimension and also spins $1/2$ and $2$ in four dimensions. Subsequently these methods were used to
find the antisymmetric tensor propagator in dS \cite{PVA} and also the propagators of various $p-$forms of interest in
supergravity/string theory in AdS \cite{IB, IB1, BNV}.\footnote{The basis of bitensor structures used in the AdS/CFT
literature differs from the original basis of \cite{AJ}. We comment more on that in a subsequent footnote.} However,
only quite recently was this method extended to spin $1/2$ field in arbitrary dimension \cite{WM}, and the spin $3/2$
field has not been treated so far. \footnote{Aspects of spin-3/2 propagation were studied already by
Lichnerowicz~\cite{Lich61}; more recent interesting studies on higher spin fields in constant curvature spaces 
(\eg the discovery of partial masslessness and new local gauge invariances) can be found in \cite{DW} and references therein.} 

Since dS is not a supersymmetric background, it may seem uninteresting to consider
the superpartner of the graviton in it. However, if dS is to be reconciled with the current lore of a fundamental
theory, i.e. string theory, then the lack of supersymmetry in de Sitter space should be understood as a symmetry which
is present in the theory but broken in the specific vacuum state. Given that superymmetry breaking in supergravity
leads to massive gravitinos, massive spin $3/2$ fields are essential for understanding the effective description of quantum 
gravity processes in de Sitter space. Motivated by this, we find in this paper the propagator of massive spin
$3/2$ in an arbitrary dimension. We use this result to test the dS/CFT correspondence. 

Much of the literature on dS/CFT is concerned with scalar fields in dS and the prescription for correlators of their
dual operators \cite{AS, BMS, SV}. Only very recently the conformal dimensions of the dual operators for massive spin
$1$ and spin $2$ fields were found \cite{OCo}. We use the previously known spin $1/2$ propagator and the newly found
spin $3/2$ propagator to calculate the conformal dimensions of the dual operators. In both cases we find, in agreement
with analytic continuation, that it is always complex. Namely we find $\Delta = \frac{n-1}{2} + i m$, where $n$ is the
dimension of space-time and $m$ is the mass parameter. This, of course, implies that the conjectured dual description
would {\it always} be nonunitary, unlike in the bosonic case, where there is some limited range of real conformal
dimensions for every spin in every dimension of space time.\footnote{It should be noted that the work~\cite{GKS} gives
a proof that the SO$(d,1)$ isometry group of dS cannot have unitary representations living on dS, while our results
indicate that the {\em dual} CFT should be non-unitary.} There has already been an indication that the current
formulation of the dS/CFT correspondence may be problematic \cite{DLS}, although several works~\cite{BBM, KV} have
proposed ways out of this difficulty. In our opinion the lack of unitary dual for fermions is a strong evidence that
the correspondence should be reformulated. 

This paper is organized as follows. In Section 2 we review the invariant bitensors in maximally symmetric spaces and their properties. In Section 3 we derive the massive spin $3/2$ propagator in arbitrary dimension in a maximally symmetric space. In Section 4 we find the conformal dimensions of the dual operators for massive spin $1/2$ and spin $3/2$ in dS/CFT.

\section{Background Material}

\label{sec:background}

In this section we review the necessary background, first given in~\cite{AJ, AL, WM}, that we will use in
the next section to find the spin $3/2$ propagator in terms of intrinsic geometric objects.\footnote{While
our treatment is based on Refs.~\cite{AJ, AL, WM}, the idea of constructing propagators from intrinsic
bitensors dates at least back to Lichnerowicz~\cite{Lich61}.}

We will consider a maximally symmetric space of dimension $n$. The geometric objects of interest are the geodesic
distance $\mu (x, x^{\prime})$ between two points $x$ and $x^{\prime}$, the unit tangent vectors $n_{\sigma} (x,
x^{\prime})$ and $n_{\sigma^{\prime}} (x, x^{\prime})$ to the geodesic at $x$ and at $x^{\prime}$, respectively, the
parallel propagator of vector indices $g^{\mu}{}_{\nu^{\prime}} (x, x^{\prime})$, and the parallel propagator of spinor
indices $\Lambda^{\alpha}{}_{\beta^{\prime}} (x, x^{\prime})$.\footnote{Primed indices will always refer to the point
$x^{\prime}$ and unprimed ones to $x$.} Let us recall the definition of all these maximally symmetric bitensors. 
The vectors $n_\sigma, n_{\sigma'}$ are defined by 
\be 
  n_{\sigma} = \nabla_{\sigma} \mu (x, x^{\prime}) \qquad {\rm and} \qquad n_{\sigma^{\prime}} = \nabla_{\sigma^{\prime}} \mu (x, x^{\prime}) \, , 
\ee 
where $\nabla_{\sigma}$ is the covariant derivative. It is important to note that 
\be 
  n_{\sigma} = - g_{\sigma}{}^{\rho^{\prime}} n_{\rho^{\prime}} \, . 
\ee 
The parallel propagators of vector and spinor indices satisfy by definition 
\be 
  V^{\mu} (x) = g^{\mu}{}_{\nu^{\prime}} (x, x^{\prime}) V^{\nu^{\prime}} (x^{\prime}) \ , \qquad \Psi^{\alpha} (x) = \Lambda^{\alpha}{}_{\beta^{\prime}} (x, x^{\prime}) \Psi^{\beta^{\prime}} (x^{\prime}) 
\ee 
for every parallel-transported vector $V^{\mu}(x)$ and spinor $\Psi^{\alpha} (x)$, respectively.

The covariant derivatives of the above bitensors, essential for our calculations, are given by
\bea
  \nabla_{\mu} n_{\nu} &=& A \, (g_{\mu \nu} - n_{\mu} n_{\nu}) \nn \ , \\
  \nabla_{\mu^{\prime}} n_{\nu} &=& C \, (g_{\mu^{\prime} \nu} + n_{\mu^{\prime}} n_{\nu}) \nn \ ,\\
  \nabla_{\mu} g_{\nu \rho^{\prime}} &=& - (A + C) \, (g_{\mu \nu} n_{\rho^{\prime}} + g_{\mu \rho^{\prime}} n_{\nu}) \nn \ , \\
  \nabla_{\mu} \Lambda^{\alpha}{}_{\beta^{\prime}} &=& \frac{1}{2} (A + C) \, [\,(\Gamma_{\mu} \Gamma^{\nu} n_{\nu} -
       n_{\mu}) \, \Lambda]^{\,\alpha}{}_{\beta^{\prime}} \nn \ ,\\ 
  \nabla_{\mu^{\prime}} \Lambda^{\alpha}{}_{\beta^{\prime}}
       &=& - \frac{1}{2} (A + C) \, [\,(\Gamma_{\mu^{\prime}} \Gamma^{\nu^{\prime}} n_{\nu^{\prime}} - n_{\mu^{\prime}}) \,
  \Lambda]^{\,\alpha}{}_{\beta^{\prime}} \, , \label{CovDer}
\eea
where $A$ and $C$ are the following functions of the geodesic distance:
\bea
&&{\rm for \,\,\,\, \R^{n}:} \qquad A (\mu) =  \frac{1}{\mu} \, , \qquad C (\mu) = - \frac{1}{\mu} \nn \, , \\
&&{\rm for \,\,\,\, dS:} \qquad A (\mu) =  \frac{1}{R} \cot \frac{\mu}{R} \, , \qquad C (\mu) = - \frac{1}{R \sin (\frac{\mu}{R})}\ , \nn \\
&&{\rm for \,\,\,\, AdS:} \qquad A (\mu) =  \frac{1}{R} \coth \frac{\mu}{R} \, , \qquad C (\mu) = - \frac{1}{R \sinh (\frac{\mu}{R})} \, ,
\label{AC} 
\eea
$R$ being the radius of (A)dS space. The covariant gamma matrices in (\ref{CovDer}) satisfy the usual relation $\{\Gamma^{\mu}, \Gamma^{\nu}\} = 2 g^{\mu \nu}$.

In $\R^n, \S^n, \H^n$ every two points are connected by a geodesic, but in pseudo-Riemannian spaces this is not the case: a notable counter example is, for instance, the pair $x, x^{\prime}$ in de Sitter space such that the geodesic between $x$ and the antipodal point of $x^{\prime}$ is timelike \cite{SSV}.  Thus the geodesic distance is not a globally defined quantity in the cases of physical interest. This problem can be circumvented in the following way. Let the $n$-dimensional maximally symmetric space be defined as the set of points in $\R^{n+1}$ satisfying:
\begin{equation}
X^a X^b \eta_{a b} = R^2 \, , \label{embed}
\end{equation}
where $\eta$ is a flat metric with the appropriate signature. Then for two points $x$ and $x^{\prime}$ connected by a geodesic:
\begin{equation}
\cos \left( \frac{\mu (x, x^{\prime})}{R} \right) = \frac{X^a (x) X^b (x^{\prime}) \eta_{a b}}{R^2} \, .
\end{equation}
Now, introducing the variable $z = \cos^2 \frac{\mu}{2 R}$, one obtains:
\begin{equation}
z = \frac{1}{2} \left( 1 + \frac{X^a (x) X^b (x^{\prime}) \eta_{a b}}{R^2} \right) \, . \label{zi}
\end{equation}
Clearly~(\ref{zi}) gives a globaly defined quantity that can be viewed as a "generalization" of $\mu$.

\section{Massive spin 3/2 propagator}

Now we turn to the propagator of the massive spin 3/2 field. Let us denote the gravitino field by $\Psi^{\alpha}_{\lambda} (x)$. In a maximally symmetric state $|\, s >$ the propagator is $ S^{\alpha \beta^{\prime}}_{\lambda \nu^{\prime}} (x, x^{\prime}) = <s\,| \Psi^{\alpha}_{\lambda} (x) \Psi^{\beta^{\prime}}_{\nu^{\prime}} (x^{\prime}) |\,s>$. The field equations imply that $S$ satisfies 
\be
(\Gamma^{\mu \rho \lambda} D_{\rho} - m \, \Gamma^{\mu \lambda})^{\alpha}{}_{\gamma} S_{\lambda \nu^{\prime}}{}^{\gamma}{}_{\beta^{\prime}} =
\frac{\delta (x-x^{\prime})}{\sqrt{-g}} g^{\mu}{}_{\nu^{\prime}} \, \delta^{\alpha}{}_{\beta^{\prime}} \label{EoM} 
\ee
For a massive spin $3/2$ and a vanishing source the above field equation implies two constraints (for a very clear recent explanation of this point, see~\cite{GPvN}): 
\be
\Gamma^{\lambda} S_{\lambda \nu^{\prime}} = 0 \, ,\qquad \,\,\,\,\, D_{\lambda} \, S^{\lambda}{}_{\nu^{\prime}} = 0 \label{constr} \, ,
\ee
where we have suppressed the spinor indices for brevity. Using (\ref{constr}) and the fact that
\be
\Gamma^{\mu \nu \lambda} = \Gamma^{\mu} \Gamma^{\nu} \Gamma^{\lambda} - g^{\mu \nu} \Gamma^{\lambda} - g^{\nu \lambda} \Gamma^{\mu} + g^{\mu\lambda} \Gamma^{\nu} \, , 
\ee
(\ref{EoM}) becomes
\be
\Gamma^{\rho} D_{\rho} S_{\lambda \nu^{\prime}} + m \, S_{\lambda \nu^{\prime}} = 0 \label{sEoM} \, .
\ee
For the time being we have set the source to zero, as it complicates the field equations due to its appearance in the
constraints~(\ref{constr}). Hence we will be solving the homogeneous system of equations and the role of the
$\delta$-function source term will be in imposing the appropriate boundary conditions at the very end of the
computation. Namely we will impose that the solution is singular  at $\mu=0$ and that the strength of the singularity is the same as the one in flat space.\footnote{Solutions we find are automatically convergent at
$\infty$.} This is essentially the strategy followed in all original
literature on obtaining of propagators in maximally symmetric spaces in terms of intrinsic geometric objects \cite{AJ,
AL, WM}.  \newline 
\indent
The most general Ansatz for the spin 3/2 propagator has 10 tensor structures and we write it in the following form:
\bea
  S_{\lambda \nu^{\prime}}{}^{\alpha}{}_{\beta^{\prime}} 
    &=& \alpha (\mu) \, g_{\lambda \nu^{\prime}} \Lambda^{\alpha}{}_{\beta^{\prime}} + 
        \beta (\mu) \, n_{\lambda} n_{\nu^{\prime}} \Lambda^{\alpha}{}_{\beta^{\prime}} + 
        \gamma (\mu) \, g_{\lambda \nu^{\prime}} (n_{\sigma} \Gamma^{\sigma} \Lambda)^{\alpha}{}_{\beta^{\prime}} \nn \\ 
    && + \delta (\mu) \, n_{\lambda} n_{\nu^{\prime}} (n_{\sigma} \Gamma^{\sigma} \Lambda)^{\alpha}{}_{\beta^{\prime}} + 
        \varepsilon (\mu) \, n_{\lambda} (\Gamma_{\nu^{\prime}} \Lambda)^{\alpha}{}_{\beta^{\prime}} + 
        \theta (\mu) \, n_{\nu^{\prime}} (\Gamma_{\lambda} \Lambda)^{\alpha}{}_{\beta^{\prime}} \nn \\ 
    && + \tau (\mu) \, n_{\lambda} (n_{\sigma} \Gamma^{\sigma} \Gamma_{\nu^{\prime}} \Lambda)^{\alpha}{}_{\beta^{\prime}} 
       + \omega(\mu)\, n_{\nu^{\prime}} (n_{\sigma} \Gamma^{\sigma} \Gamma_{\lambda} \Lambda)^{\alpha}{}_{\beta^{\prime}} \nn \\ 
    && + \pi (\mu) \, (\Gamma_{\lambda} \Gamma_{\nu^{\prime}} \Lambda)^{\alpha}{}_{\beta^{\prime}}  
       + \kappa (\mu)\, (n_{\sigma} \Gamma^{\sigma} \Gamma_{\lambda} \Gamma_{\nu^{\prime}} \Lambda)^{\alpha}{}_{\beta^{\prime}} 
          \label{ansatz} 
\eea
Substituting it in (\ref{sEoM}) and using (\ref{CovDer}), a rather tedious but straightforward calculation gives a
system of $10$ equations for the $10$ coefficient functions $\alpha, ..., \kappa$ in (\ref{ansatz}). It may seem that
it does not matter what basis of tensors one would choose for the Ansatz. Namely one could take the term $(n\cdot
\Gamma)$ to be on the right of every term containing other gamma matrices instead of on the left as in (\ref{ansatz}).
Or one could take every term in the Ansatz to be symmetric or antisymmetric in the $(\lambda, \nu^{\prime})$ indices,
i.e. for example instead of the $\varepsilon, \theta$ terms in (\ref{ansatz}) one could write $\varepsilon (\mu) n_{(
\lambda} \Gamma_{\nu^{\prime})} + \theta (\mu) n_{[ \lambda} \Gamma_{\nu^{\prime} ]}$. Of course, all these different
choices amount to merely rearranging of the terms in the propagator. However, our experience showed that~(\ref{ansatz})
gives the simplest equations. All other Ans\"atze result in equations which are significantly more complicated and more
difficult to deal with.\footnote{In the literature on AdS/CFT correspondence the widely used basis is a different one,
based on derivatives of the chordal distance variable $u$ in AdS \cite{Freed}. Although it is useful, for example,  for
decoupling of physical and gauge parts of gauge boson propagators in AdS, for our purposes the original basis of
\cite{AJ} is the more convenient one. The two bases are related by a linear transformation given in \cite{Freed}, whose
coefficients are scalar functions of $u$.} Denoting $\frac{d}{d \mu} \equiv {}^{\prime}$, we write out the system of
equations that (\ref{sEoM}) gives for the coefficient functions in (\ref{ansatz}): 
\be
\gamma^{\prime} + \frac{1}{2} (A-C) (n-1) \gamma + 2 C \theta + m \alpha = 0 \label{eq1}
\ee
\be
\delta^{\prime} + \frac{1}{2} (A - C) (n+1) \delta  - (A + C) (n-2) \theta + m \beta = 0 \label{eq2}
\ee
\be
\alpha^{\prime} + \frac{1}{2} (A + C) (n-1) \alpha - 2 C \omega + m \gamma = 0 \label{eq3}
\ee
\be
\beta^{\prime} - (A - C) \beta  + \frac{1}{2} (A + C) (n-1) \beta + (A  + C) (n-2) \omega + m \delta = 0 \label{eq4}
\ee
\be
\!\,\,\, - (A + C) \alpha + C \beta + \tau^{\prime} + [\frac{1}{2} (A - C) (n-1) + A] \tau  - (A + C) (n-2) \pi + m \varepsilon = 0
\ee
\be
- (A + C) \alpha + A \beta + \omega^{\prime} + [\frac{1}{2} (A - C) (n-1) + A] \omega + m \theta = 0
\ee
\be
\!\,\,\, (A + C) \gamma - C \delta + \varepsilon^{\prime} + [\frac{1}{2} (A + C) (n-1) - A] \varepsilon + (A + C) (n-2) \kappa + m \tau = 0
\ee
\be
(A + C) \gamma - A \delta + \theta^{\prime} + [\frac{1}{2} (A + C) (n-1) - A] \theta + m \omega = 0
\ee
\be
A \varepsilon - C \theta + \kappa^{\prime} + [2 A + \frac{1}{2} (A - C) (n-3)] \kappa + m \pi = 0 \label{eq9}
\ee
\be
- A \tau + C \omega + \pi^{\prime} + \frac{1}{2} (A + C) (n-3) \pi + m \kappa = 0 \label{eq10} \, .
\ee
\newline
\noindent
The constraint $(\Gamma \cdot S)_{\nu^{\prime}} = 0$ in (\ref{constr}) gives the following equations:
\bea
\alpha + \tau + n \pi &=& 0 \nn \\
\beta - (n - 2) \omega &=& 0 \nn \\
- 2 \gamma + \delta + n \theta &=& 0 \nn \\
- \gamma + \varepsilon - (n - 2) \kappa &=& 0 \label{C1} \, .
\eea
Finally, the constraint $(D \cdot S)_{\nu^{\prime}} = 0$ in~(\ref{constr}) implies \\ 
\bea
&&\!\!\!\!- \alpha^{\prime} - (A + C) \left( n - \frac{1}{2} \right) \alpha + \beta^{\prime} + A (n-1) \beta +
\omega^{\prime} + \left[C + \frac{1}{2} (3 A + C) (n-1)\right] \omega = 0 \nn \\ \nn \\
&&\qquad \frac{1}{2} (A - C) \gamma + \kappa^{\prime} + \frac{1}{2} (3 A + C) (n-1) \kappa + \varepsilon^{\prime} + A (n-1) \varepsilon + C \theta = 0 \nn \\ \nn \\
&&\!\!\!\!- \gamma^{\prime} + \left[A - (A + C) \left(n-\frac{1}{2}\right)\right] \gamma + \delta^{\prime} + A (n-1)
\delta + \theta^{\prime} + \left[C + \frac{1}{2}(A + C) (n-1)\right] \theta = 0 \nn \\ \nn \\
&&\qquad - \frac{1}{2} (A + C) \alpha + \pi^{\prime} + \frac{1}{2} (A + C) (n-1) \pi + \tau^{\prime} + A (n-1) \tau + C \omega = 0 \label{Cd4}
\eea
\newline
\indent
Although it would have been a rather daunting task to solve the coupled
equations resulting from (\ref{EoM}), the equivalent system of field equations
(\ref{sEoM}) and constraints (\ref{constr}), giving (\ref{eq1}-\ref{Cd4}), can
be easily reduced to a single second order ordinary differential equation in
the following way. We view the functions $\alpha, ..., \kappa$ and their
derivatives $\alpha^{\prime}, ..., \kappa^{\prime}$ as independent variables.
Setting aside 2 field equations, for example (\ref{eq1}) and (\ref{eq3}), and
treating the functions $\alpha, \alpha^{\prime}, \gamma, \gamma^{\prime}$ as
known parameters, we are left with an algebraic system of $16$ equations ($8$
field equation and $8$ constraints) for $16$ ``variables'' $\beta,
\beta^{\prime}, \delta, \delta^{\prime}, ..., \kappa, \kappa^{\prime}$. This
system has a unique solution (we verified the uniqueness by confirming that the
determinant of the system of linear equations is nonzero). By substituting this
solution, namely the expressions of the rest of the variables in terms of
$\alpha, \alpha^{\prime}, \gamma, \gamma^{\prime}$, in (\ref{eq1}) and
(\ref{eq3}) we obtain two first order ODEs, which are easily reduced to a
single second order ODE, as we will see later. 

However, once the coefficient functions and their derivatives are viewed as
independent variables, $4$ additional constraints appear, namely the equations
obtained by taking $\frac{d}{d \mu}$ of (\ref{C1}). One can check that these
$4$ equations are linear combinations of the $10$ field equations
(\ref{eq1}-\ref{eq10}) and $8$ constraints (\ref{C1}-\ref{Cd4}), so we learn
nothing new from them. However, they are much simpler than the $4$ constraints
(\ref{Cd4}), making it calculationaly preferable to keep the former four
equations and drop instead (\ref{Cd4}), which are also linear combinations of
(\ref{eq1}-\ref{eq10}), (\ref{C1}) and  $\frac{d}{d \mu}$(\ref{C1}). Of course,
dropping (\ref{Cd4}) as opposed to dropping $\frac{d}{d \mu}$(\ref{C1})
produces different algebraic answers, because after setting aside $2$ field
equations the linear dependency of $4$ constraints on the remaining $8$ field
equations and $8$ constraints is lost; the answers coincide for solutions of
the two field equations that we have set aside to solve as differential
equations. In other words, the different algebaric answers differ from each
other only  by terms proportional to the two field equations set aside.
Whichever algebraic answer one takes may look (and in fact it does look)
horrible, because of many terms vanishing for the solutions of the two
differential equations that are set aside. Hence the algebraic outcome of the
procedure outlined in the previous paragraph should be simplified by imposing
on it the two remaining differential equations as algebraic constraints. This
is a key observation which makes the calculation feasible; without it the
algebraic answers look utterly intractable.

Before giving the algebraic answer for eight of the coefficient functions in terms of the remaining two\footnote{Due to the simplification procedure explained in the previous paragraph these expressions do not contain derivatives of the two functions.}, let us make one more observation. The field equations (\ref{eq1}-\ref{eq10}) obviously split into five 'conjugate' couples of the form:
\be
x^{\prime} + ... + m y = 0 \, ,\qquad y^{\prime} + ... + m x = 0 \, ,
\ee
where $(x,y)$ is $(\alpha, \gamma)$, $(\beta, \delta)$, $(\omega, \theta)$, $(\varepsilon, \tau)$ or $(\pi, \kappa)$. Hence there are five natural options for the choice of the couple of equations to be solved as differential at the end of the algebraic procedure. It may seem that it is equally easy (or equally difficult) to choose any couple, but we have found that the simplest second order differential equation is obtained for the $(\pi, \kappa)$ pair\footnote{This is actually not true for flat space where the simplest differential equations are obtained for the $(\beta, \delta)$ pair. We will come back to this in Subsection 3.1.}. Hence we write down the algebraic solutions for $\alpha, \beta, \gamma, \delta, \varepsilon, \theta, \tau, \omega$ in terms of $\kappa$ and $\pi$:
\bea \label{alg}
\omega &=& \frac{4mC(n-2) \kappa + ((A+C)^2(n-2)^2-4m^2) \pi}{(n-2)^2/R^2 + 4 m^2} \, ,\nn \\ \nn \\
\theta &=& \frac{((A-C)^2(n-2)^2-4m^2) \kappa - 4mC(n-2) \pi}{(n-2)^2/R^2 + 4 m^2} \, , \\ \nn \\
\tau &=& \frac{4mC(n-2) \kappa + ((A+C)^2(n-2)^2-4m^2) \pi}{(n-2)^2/R^2 + 4 m^2} \, ,\nn \\ \nn \\
\varepsilon &=& \frac{([-(A-C)^2 - 2/R^2] (n-2)^2 - 4m^2) \kappa + 4mC(n-2) \pi}{(n-2)^2/R^2 + 4 m^2} \, , \nn \\ \nn \\
\alpha = - \tau - n\pi \, , \qquad &\beta& = (n-2) \omega \, , \qquad \gamma = \varepsilon - (n-2) \kappa \, , \qquad \delta = 2\varepsilon + 2 (n-2) \kappa -n\theta \, , \nn
\eea
where we have used the relation $C^2 - A^2 = 1/R^2$. Further, from (\ref{alg}) we can immediately see that
\be
\tau = \omega \qquad {\rm and} \qquad \varepsilon + \theta = - 2 \kappa \, . \label{Sym}
\ee
In view of (\ref{Sym}) the Ansatz (\ref{ansatz}) becomes:
\bea
S_{\lambda \nu^{\prime}}{}^{\alpha}{}_{\beta^{\prime}} &=& \alpha (\mu) \, g_{\lambda \nu^{\prime}} \Lambda^{\alpha}{}_{\beta^{\prime}} + \beta (\mu) \, n_{\lambda} n_{\nu^{\prime}} \Lambda^{\alpha}{}_{\beta^{\prime}} + \gamma (\mu) \, g_{\lambda \nu^{\prime}} (n_{\sigma} \Gamma^{\sigma} \Lambda)^{\alpha}{}_{\beta^{\prime}} \nn \\ &+& \delta (\mu) \, n_{\lambda} n_{\nu^{\prime}} (n_{\sigma} \Gamma^{\sigma} \Lambda)^{\alpha}{}_{\beta^{\prime}} + 2 \, \theta (\mu) \, (\Gamma_{[\lambda} n_{\nu^{\prime}]} \Lambda)^{\alpha}{}_{\beta^{\prime}} \nn \\ &+& 2 \, \omega (\mu) \, (n_{\sigma} \Gamma^{\sigma} n_{(\lambda} \Gamma_{\nu^{\prime})} \Lambda)^{\alpha}{}_{\beta^{\prime}} \nn \\ &+& \pi (\mu) \, (\Gamma_{\lambda} \Gamma_{\nu^{\prime}} \Lambda)^{\alpha}{}_{\beta^{\prime}} + \kappa (\mu) \, ( \Gamma_{[\lambda} (n \cdot \Gamma) \Gamma_{\nu^{\prime}]} \Lambda)^{\alpha}{}_{\beta^{\prime}} \, .
\eea
This may seem to suggest that the natural Ansatz to start with is the one in which every term is symmetric or antisymmetric in $(\lambda, \nu^{\prime})$. But we want to stress again that, although counter intuitive, such a starting point produces much more complicated equations and constraints than (\ref{eq1}-\ref{Cd4}).

Using~(\ref{Sym}) the differential equations for $\kappa$ and $\pi$, (\ref{eq9}) and (\ref{eq10}), acquire the form:
\bea \label{kp}
-(A+C) \theta + \kappa^{\prime} + \frac{1}{2} (A-C) (n-3) \kappa + m \pi &=& 0 \nn \\
(C-A) \omega + \pi^{\prime} + \frac{1}{2} (A+C) (n-3) \pi + m \kappa &=& 0 \, ,
\eea
where $\theta$ and $\omega$ are given in (\ref{alg}). Clearly one can solve algebraically the second equation for $\kappa$. Differentiating the result one obtains also $\kappa^{\prime}$ in terms of $\pi$, $\pi^{\prime}$ and $\pi^{\prime \prime}$. Substituting these in the first equation gives a second order ODE for $\pi (\mu)$. We will consider it both for the cases of dS and AdS spaces in turn. To start with, however, we will consider the flat space case, which can be used to fix the normalization.

\subsection{Minkowski space}

The system of equations (\ref{eq1}-\ref{eq10}) simplifies significantly for flat space due to $A+C = 0$ (see (\ref{AC})). One can immediately see that the most convenient pair of differential equations to solve is the pair (\ref{eq2}) and (\ref{eq4}) for $\beta$ and $\delta$, unlike in the curved space case where (as we already mentioned) the natural pair is $\kappa, \pi$.

The equations of motion are 
\bea \label{BD}
&&\delta^{\prime} +\frac{1}{\mu} (n+1) \delta + m \beta = 0 \nn \\
&&\beta^{\prime}-\frac{2}{\mu} \beta + m \delta = 0 \, .
\eea
Using the second equation to express $\delta, \delta^{\prime}$ in terms of $\beta$, $\beta^{\prime}$ and $\beta^{\prime \prime}$ and substituting these in the first we obtain:
\be
\beta^{\prime \prime} + \frac{n-1}{\mu} \beta^{\prime} - \left(\frac{2n}{\mu^2}+m^2\right) \beta = 0 \, . \label{eq:BetaEOM}
\ee
The solution is
\be
\beta (\mu) = \CC (m \mu)^{1-n/2} K_{n/2 + 1} (m \mu) \, , \label{FlatSol}
\ee
where $\CC$ is a constant and we have retained only the solution of the Bessel equation which diverges for $\mu \rightarrow 0$. 

To fix $\CC$ we have to reinsert the source in (\ref{EoM}) and see what does it
give rise to in the constraints and field equations (\ref{eq1}-\ref{Cd4}).
Doing so carefully,\footnote{The calculation essentially amounts to repeating
the one done e.g. in~\cite{GPvN}, but in arbitrary dimension and with a
non-conserved source: in flat space $\nabla_{\rho} J^{\rho}{}_{\nu^{\prime}} =
n_{\nu^{\prime}} \delta^{\prime} (x - x^{\prime})$.} one finds that the zero on
the righthand side of the first equation in (\ref{BD}) changes to
\be
\frac{5 (n-2)}{m^2 (n-1)} \left( \frac{\delta^{\prime} (x - x^{\prime})}{\mu} - \delta^{\prime \prime} (x - x^{\prime}) \right) \, .
\ee
Denoting the numerical factor $\frac{5 (n-2)}{m^2 (n-1)}$ as $a$, Eq.~(\ref{eq:BetaEOM}) becomes
\be
  \beta^{\prime \prime} + \frac{n-1}{\mu} \beta^{\prime} - \left(\frac{2n}{\mu^2}+m^2\right) \beta = a \left( \frac{\delta^{\prime} (x - x^{\prime})}{\mu} - \delta^{\prime \prime} (x - x^{\prime}) \right). \label{eq:BetaEOMWSrc}
\ee
This equation is understood in the sense of distributions.
To fix the normalization constant, we multiply~(\ref{eq:BetaEOMWSrc}) by
$\mu^2$ and use $\beta^{\prime \prime} + \frac{(n-1)}{\mu} \beta^{\prime} = \Box
\beta(\mu)$, giving
\bea
  \mu^2 \Box \beta - \left(2n+m^2 \mu^2 \right) \beta 
     &=& a \left( \mu \delta^{\prime} (x - x^{\prime}) - \mu^2\delta^{\prime \prime} (x - x^{\prime}) \right) \nonumber \\
     &=& -n(n+2) a \delta (x - x^{\prime}). \label{eq:betaEOMWSrc2}
\eea
We now multiply~(\ref{eq:betaEOMWSrc2}) by a test function $f(\mu)$ that is regular and non-zero at $\mu=0$, and
integrate the resulting equation over a ball of radius $\epsilon \ra 0$ centered at $\mu=0$. 
Using the following normalization of the Neumann function,
\be 
  K_{\frac{n}{2}+1}(x) \ra 2^\halfof{n} \Gamma\left(\halfof{n}+1\right) x^{-\halfof{n}-1} \ \ \ \ \ \mbox{for} \ x \ra 0 \, , \label{eq:KNorm}
\ee
we obtain
\be
  \CC = m^{n+1} \frac{na}{(2\pi)^\halfof{n} n} =
       m^{n-1} \frac{5(n-2)}{(2\pi)^\halfof{n} (n-1)} \, . \label{eq:BetaNorm}
\ee 
Having found $\beta$, one can find $\delta$ from the second equation in
(\ref{BD}) and the rest of the coefficient functions are expressed in terms of
$\beta$ and $\delta$ in the following way: 
\bea \label{flat}
&&\tau = \omega = \frac{\beta}{n-2} \, , \qquad \pi = - \left( \frac{1}{n-2} +\frac{n+2}{m^2 \mu^2} \right) \beta +\frac{1}{m \mu} \delta \, , \nn \\ \nn \\
&&\varepsilon = \frac{1}{n-2} \left( - \frac{4}{m \mu} \beta + \delta \right) \, , \qquad \theta = \frac{1}{n-2} \left( - \frac{2n}{m \mu} \beta + \delta \right) \, , \nn \\ \nn \\
&&\kappa = - \frac{1}{2} (\varepsilon + \theta) \, , \qquad \alpha = - \tau - n \pi \, , \qquad \gamma = \frac{1}{2} (\delta + n \theta) \, .
\eea
Thus, as in the case of other spins, the flat space result is expressed in terms of Bessel functions. We would also like 
to mention that as the flat space solution suggests, our considerations here (and below for curved space) are strictly 
valid for the massive case. The massless limit requires a special treatment, as done for spin $1$ back in \cite{AJ}, 
and we will not do that in this paper.\footnote{We remind the reader that the constraints we used to reduce the problem 
of solving ten coupled first order differential equations to solving only two are not present for $m=0$.}

\subsection{dS space}

Now we are going back to the system (\ref{kp}). Substituting $A$ and $C$ from
(\ref{AC}) and changing to the globally defined variable $z = \cos^2
\frac{\mu}{2R}$ (see Section 2), we obtain the following equation for $\pi$: 
\bea \label{FE}
&&\left\{\left[(n-2)^2 + 4 m^2 R^2\right] z^4 +(6n - 2n^2 -4 -8 m^2 R^2) z^3
+(n^2-2n+4m^2R^2)z^2\right\} \frac{d^2}{dz^2} \pi + \nn \\
&&\left\{\left[(n-2)^2 + 4m^2R^2\right]n z^3 + \left[-6m^2R^2n+4(n-1)^2
-\frac{3}{2}n^3\right]z^2 \right. + \nn \\
&&\left.\left(\frac{n^2}{2}-n +2m^2R^2\right)nz\right\} \frac{d}{dz} \pi + \nn \\
&&\left\{ \left(5m^2+\frac{n^4}{4}+4m^4-6m^2n+1+\frac{13}{4}n^2-3n
-\frac{3}{2}n^3+2n^2m^2R^2\right) z^2 \right. + \nn \\
&&\left(-4m^4-6m^2+\frac{11}{2}n
+5m^2n-3+\frac{5}{4}n^3-\frac{1}{4}n^4-\frac{7}{2}n^2-2n^2m^2R^2\right) z  \nn \\
&&\left. -m^2R^2(n-1)+\frac{n}{4}\left(3n-n^2-2\right)\right\} \pi = 0
\eea
Making the substitution $\pi (z) = \sqrt{z} \, \tilde{\pi} (z)$, (\ref{FE}) becomes:
\bea \label{FEp}
&&\left\{\left[(n-2)^2 +
4m^2R^2\right]z^2(z-1)-\left[n(n-2)+4m^2R^2\right]z(z-1)\right\} \frac{d^2}{dz^2} \tilde{\pi} + \nn \\
&&\left\{\left[n^2(n-3)+4+4m^2R^2(n+1)\right]z^2+\left[-\frac{3
n^3}{2}-2m^2R^2(4+3n)+2n(n+1)\right] z \right.+ \nn \\
&&\left. \frac{n^3}{2}-2n+2m^2R^2(n+2) \right\} \frac{d}{dz} \tilde{\pi} + \nn
\\
&&\left\{\left[4m^4R^4 +\frac{n^4}{4} -n^3 +4m^2R^2+n^2-4m^2R^2n+2n^2m^2R^2
\right] z \right.  \nn \\ 
&&\left.
-\frac{n^4}{4}+\frac{n^3}{2}-n^2-4m^4R^4-2n^2m^2R^2-4m^2R^2+2m^2R^2n+2n
\right\} \tilde{\pi} = 0 \, .\nn \\
\eea
This equation is of the type:
\bea \label{Heun}
&&z(z-1)(z-a)y^{\prime \prime}(z) + \left\{ (b+c+1)z^2 -
\left[b+c+1+a(d+e)-e\right]z +ad
\right\} y^{\prime} (z) \nn \\
&&+ (bc\, z-q) y(z) = 0 \, ,
\eea
which is known as Heun's equation~\cite{PZ, Heun}. Its solutions, denoted by $F(a,q;b,c,d,e;z)$, have
in general 4 singular points $z_0=0,1,a,\infty$. Near each singularity the function behaves as a combination of two
terms that are powers of $(z-z_0)$ with the following exponents:
$\{0, 1-d\}$ for $z_0 = 0$, $\{0, 1-e\}$ for $z_0=1$, $\{0, d+e-b-c\}$ for $z_0=a$, and $\{b,c\}$ (that is, 
$z^{-b}$ or $z^{-c}$) for $z\ra \infty$. 

Comparing (\ref{FEp}) and (\ref{Heun}), we can read off the parameters for our case:
\bea \label{param}
&&a = \frac{n(n-2)+4m^2R^2}{(n-2)^2 + 4m^2R^2} \, , \qquad b = \frac{n}{2}+imR \, , \qquad c = \frac{n}{2}-imR \, ,
\qquad d = e = \halfof{n}+1 \nn \\ \nn \\
&& q = \frac{-\frac{n^4}{4}+\frac{n^3}{2}-n^2-4m^4R^4-2n^2m^2R^2-4m^2R^2+2m^2R^2n+2n}{(n-2)^2 + 4m^2R^2} \, .
\eea
The propagator is the particular Heun's function whose singularity at $z=1$ (which corresponds 
to $\mu=0$) is of the same type as that in flat space, namely $z^{-\halfof{n}} \propto \mu^{-n}$, which singles out the
exponent $1-e = -\halfof{n}$. This same argument can also be used to fix the normalization of the propagator, by
comparing it with~(\ref{eq:BetaNorm}) via~(\ref{flat}).  

An important property of the propagator for our purposes is its behaviour for $z\ra \infty$ ($z\ra \infty$
corresponds to going to the boundary of dS). As already mentioned, for $z\ra\infty$ Heun's functions behave as a combination of
$z^{-b}$ and $z^{-c}$, which in view of~(\ref{param}) becomes $z^{-\halfof{n}+imR}$ and $z^{-\halfof{n}-imR}$.

The singularity at $z=a$ turns out to be harmless: the exponents at $z=a$ are (using~(\ref{param}) again) $\{0,2\}$; namely, 
the propagator is actually regular at $z=a$. Also,
there is a singularity of the type $z^{-\halfof{n}}$ for $z\ra 0$. This is not surprising as $z=0$ corresponds to $x, x'$ 
being antipodal points of each other. Such ``extra'' antipodal singularities have also been found in bosonic
propagators~\cite{AJ, SSV}.\footnote{While the authors of~\cite{AJ} avoid this singularity by choosing a particular
(namely the Euclidean) vacuum, according to~\cite{Heun} in the case of the
Heun's function this is possible only for certain values of $q$.}

\subsection{AdS space}

The second order ODE to be solved is the same as~(\ref{FEp}), with $R \ra iR$. But the boundary conditions change slightly: One still requires the strength of the singularity at $\mu \rightarrow 0$ to be as the one in flat space, but now the propagator is required to fall off as fast as possible at spatial infinity \cite{AJ}.

\section{Fermions in dS/CFT}

In this section we use the previously known result for the propagator of massive spin $1/2$ and the massive spin $3/2$ propagator that we found in the previous section in order to test dS/CFT correspondence. Namely we will find the conformal dimension of the dual operators for both cases. The advocates of dS/CFT being merely an analytic continuation from AdS/CFT may view this as a check on our result for the spin $3/2$ propagator. We start with the simpler case of spin $1/2$ and treat it in detail; it turns out that the spin $3/2$ case is analogous because the important assymptotic properties of Heun's functions are very similar to those of the hypergeometric functions. 

In the following we will use planar coordinates, in which the metric of de
Sitter space reads
\begin{equation}
ds^2 = - dt^2 + e^{-2t} dx_i dx^i \, , \label{metric}
\end{equation}
where $i=1,..., n-1$. These coordinates cover only the half of de Sitter space which includes the past null infinity ${\cal I}^{-}$, where $t = - \infty$. This is the causal past of an observer at the south pole. Following \cite{AS}, to calculate the conformal dimension of the dual operator we need to find two things: the asymptotics for $t \rightarrow - \infty$ of the field equation, which gives the asymptotic behavior of the spinor field, and the asymptotics of the spinor propagator in dS. As the field equation for a fermion is \,\, $\slash\!\!\!\!{D} \, \Psi = m \, \Psi$ \, , we need the spin connection for the metric (\ref{metric}). Introducing the vielbeins
\begin{equation}
e^0 = dt \qquad e^i = e^{-t} dx^i \, ,
\end{equation}
we find:
\begin{equation}
\omega_i{}^{j \, 0} = e^{-t} \delta_i^j \, ,
\end{equation}
with all other components vanishing, except the ones determined by antisymmetry in the upper indices. Hence
\begin{equation}
\slash\!\!\!\!{D} = e_a^{\mu} \Gamma^a (\partial_{\mu} + \frac{1}{8} \omega_{\mu}^{b c} [\Gamma_b, \Gamma_c]) = \Gamma^0 \partial_0 + e^t \Gamma^i \partial_i -\frac{n}{2} \Gamma^0 \, .
\end{equation}
In the limit $t \rightarrow -\infty$ we obtain the field equation:
\begin{equation}
\Gamma^0 \partial_t \Psi -\frac{n}{2} \Gamma^0 \Psi = m \Psi \, .
\end{equation}
Similarly to \cite{HS}\footnote{Note that in \cite{HS}, unlike here, the signature is Euclidean.} we may set $\Gamma^0 = i$, working on the space of definite eigenvalues of $\Gamma^0$. Then the asymptotic behavior of the spinor field is
\begin{equation}
\Psi (t, \vec{x}) \rightarrow e^{(\frac{n}{2} - i m) t} \psi (\vec{x}) \, . \label{fas}
\end{equation}

The propagator for spin $1/2$ in $dS^n$ is \cite{WM}\footnote{This is the solution of the equation $[(\slash\!\!\!\!D - m)S(x, x^{\prime})]^{\alpha}{}_{\beta^{\prime}} = \delta^{\alpha}{}_{\beta^{\prime}} \delta (x - x^{\prime}) / \sqrt{-g}$. As explained in \cite{AJ} the Feynman propagator is $\lim_{\varepsilon \rightarrow 0} S(z+i \varepsilon)$. For us time ordering is not essential as we only need the limit $t, t^{\prime} \rightarrow -\infty$.}:
\begin{equation}
S(x, x^{\prime}) = [\alpha (z) + \beta (z) \, n_{\nu} \, \Gamma^{\nu}] \, \Lambda (x, x^{\prime}) \, , \label{prop}
\end{equation}
where we have suppressed spinor indices for convenience. The functions $\alpha$ and $\beta$ in (\ref{prop}) are given by:
\bea \label{CoefF}
\alpha (z) &=& \lambda \sqrt{z} F(\frac{n}{2}-i m R, \frac{n}{2} + i m R; \frac{n}{2} + 1; z) \, , \nn \\
\beta (z) &=& - \frac{\lambda}{m R} \sqrt{1-z} \left( z F(\frac{n}{2} +1 - i m R, \frac{n}{2} + 1 + i m R; \frac{n}{2} + 2; z) + \right. \nn \\
&+& \left. \frac{n}{2} F(\frac{n}{2} -i m R, \frac{n}{2} + i m R; \frac{n}{2} + 1; z) \right) \, . \label{Coef}
\eea
In (\ref{Coef}) $F$ is the hypergeometric function and $\lambda$ is a constant
given in Eq.~(28) of Ref.~\cite{WM}, which we will not write down as its
explicit form is not necessary for our purposes. For simplicity we will set $R =
1$, which is already assumed in (\ref{metric}). It can be restored easily on dimensional grounds.

Now let us calculate the asymptotics of $S^{\alpha}{}_{\beta^{\prime}} (x, x^{\prime})$ for $t \rightarrow -\infty$. Using (\ref{zi}) and the relation between the embedding coordinates $X^a$ (see (\ref{embed})) and the planar coordinates\footnote{For an excellent summary of coordinate systems and other useful properties of de Sitter space see \cite{SSV}.}:
\bea
X^0 &=& \sinh t - \frac{1}{2} e^{- t} x_i x^i \, , \nn \\
X^i &=& e^{- t} x^i, \qquad i = 1,...,n-1 \, , \nn \\
X^n &=& \cosh t - \frac{1}{2} e^{- t} x_i x^i \, ,
\eea
one finds
\be
z \rightarrow - \frac{e^{- (t + t^{\prime})}}{4} |\vec{x} - \vec{x}^{\, \prime}|^2 \, .
\ee
Since (see for ex. \cite{Er}) 
\be 
  F(a,b;c;z^{-1}) \rightarrow 1 + O(z^{-1}) \ \ \ \mbox{for} \ z \rightarrow
        \infty,
\ee 
and 
\bea \label{FhAs}
F(a,b;c;z) &=& \frac{\Gamma (c) \Gamma (b-a)}{\Gamma (b) \Gamma (c-a)} (-z)^{- a} F (a,1-c+a;1-b+a;z^{-1}) \nn \\
&+& \frac{\Gamma (c) \Gamma (a-b)}{\Gamma (a) \Gamma (c-b)} (-z)^{- b} F (b,1-c+b;1-a+b;z^{-1}) \, ,
\eea
we obtain
\be
S(x, x^{\prime}) \rightarrow \left( e^{(t+t^{\prime}) \bar{\Delta}} \frac{C_1}{|\vec{x}-\vec{x}^{\, \prime}|^{2 {\bar{\Delta}}}} + e^{(t+t^{\prime}) \Delta} \frac{C_2}{|\vec{x}-\vec{x}^{\, \prime}|^{2 {\Delta}}} \right) \frac{\Gamma \cdot (\vec{x} - \vec{x}^{\, \prime})}{|\vec{x}-\vec{x}^{\, \prime}|} \, , \label{pas}
\ee
where $C_1, C_2$ are constant coefficients, whose precise form we don't need, and $\Delta = (n-1)/2 + i m$, $\bar{\Delta} = (n-1)/2 - i m$. We have also used the fact that $\Lambda^{\alpha^{\prime}}{}_{\beta} = \delta^{\alpha^{\prime}}{}_{\beta}$ on the boundary ${\cal I}^{-}$ as this boundary is a flat $(n-1)$-dimensional space.

Now that we have all the ingredients we are ready to compute the boundary two-point correlator. Adapting the equivalent of the usual AdS/CFT prescription \cite{SG}, as proposed in \cite{AS} and used in \cite{OCo}, we have to find the coefficient of $\psi \, \psi$ (see (\ref{fas})) in the amplitude:
\be
(\Psi,\Psi) = \lim_{t \rightarrow -\infty} \int_{\cal{I}^-} d^{n-1} x \, d^{n-1} x^{\prime} \, [e^{-(n-1) t} \, \Psi (t, \vec{x}) \stackrel{\leftrightarrow}{\partial}_t S(t,\vec{x};t^{\prime}, \vec{x}^{\, \prime}) \stackrel{\leftrightarrow}{\partial}_{t^{\prime}} \Psi (t^{\prime}, \vec{x}^{\, \prime})]|_{t=t^{\prime}} \, ,
\ee
where $e^{-(n-1) t}$ is the measure factor $\sqrt{-g}$. Using (\ref{fas}) and (\ref{pas}) we obtain:
\be
(\Psi,\Psi) = \int_{\cal{I}^-} d^{n-1} x \, d^{n-1} x^{\prime} \frac{\tilde{C} \psi (\vec{x}) \psi (\vec{x}^{\, \prime})}{|\vec{x}-\vec{x}^{\, \prime}|^{2 \Delta}} \,\, \frac{\Gamma \cdot (\vec{x} - \vec{x}^{\, \prime})}{|\vec{x}-\vec{x}^{\, \prime}|} \, ,
\ee
where $\tilde{C}$ is an irrelevant constant. Hence we conclude that the two-point correlator of the dual operator ${\cal O}_{\Psi}$ is:
\be
< {\cal O}_{\Psi} (\vec{x}) {\cal O}_{\Psi} (\vec{x}^{\, \prime}) > \, = \frac{const}{|\vec{x} - \vec{x}^{\, \prime}|^{2 \Delta}} \,\, \frac{\Gamma \cdot (\vec{x} - \vec{x}^{\, \prime})}{|\vec{x}-\vec{x}^{\, \prime}|}
\ee
as appropriate for conformal invariance of operators in the spinor representation of $O(n-1)$ \cite{FGGP} with conformal dimension:
\be
\Delta = \frac{n-1}{2} + i m \, . 
\ee
As expected, the boundary correlator is consistent with conformal invariance. However, its conformal dimension is complex for every value of  $n$ and $m$, in agreement with the naive analytic continuation obtained by $m \rightarrow i m$\footnote{Recall that on dimensional grounds the mass is always multiplied by the radius $R$ and to go from AdS to dS one changes $R \rightarrow i R$.}. This is not completely surprising: even for the bosons in dS there is a range for the mass in which the conformal dimension is complex\footnote{For integral spins $s=0,1,2$, the following conformal dimensions were found in previous works: $\Delta_{s=0} = \frac{1}{2} (n-1 \pm \sqrt{(n-1)^2-4m^2})$ \cite{AS}; $\Delta_{s=1} = \frac{1}{2} (n-1 \pm \sqrt{(n-3)^2-4 m^2})$, $\Delta_{s=2} = \frac{1}{2} (n-1 \pm \sqrt{(n-1)^2-4 m^2})$ \cite{OCo}.}. In the case of bosons, as suggested in \cite{AS}, one could still avoid nonunitary dual description by assuming that in the full quantum theory of de Sitter space only particles, for which the dual theory is unitary, are stable. Clearly, for fermions there is no such option. Hence one has to either learn how to make sense of a nonunitary dual, or possibly try to modify the current formulation of the correspondence. 

The calculation of the conformal dimension of the dual operator for spin $3/2$ goes exactly 
the same way since, similarly to the hypergeometric function, the Heun's function has a behavior 
at infinity completely determined by the two exponents $b,c$:
\be
F(a,q;b,c,d,e;z) \longrightarrow Const_1 z^{-b} + Const_2 z^{-c} \qquad {\rm for} \qquad z\rightarrow \infty .
\ee
Since these exponents are the same (see (\ref{param})) as the ones that determined the behavior of the spin $1/2$ propagator (see (\ref{CoefF}) and (\ref{FhAs})) we obtain the boundary correlator\footnote{As in \cite{OCo}, one also has to use the asymptotic form of $g_{\mu \nu^{\prime}}$ for $z \rightarrow \infty$.}
\be
\frac{const}{|\vec{x} - \vec{x}^{\, \prime}|^{2 \Delta}} \,\, \frac{\Gamma \cdot (\vec{x} - \vec{x}^{\, \prime})}{|\vec{x}-\vec{x}^{\, \prime}|} \,\, \left( \delta_{ij} - \frac{2 (\vec{x}-\vec{x}^{\, \prime})_i (\vec{x}-\vec{x}^{\, \prime})_j}{ |\vec{x}-\vec{x}^{\, \prime}|^2 } \right) \label{Cor32}
\ee
with the same conformal dimension as before:
\be
\Delta_{s=3/2} = \frac{n-1}{2} + i m \, .
\ee
The form (\ref{Cor32}) agrees with the two-point correlator for spin $3/2$ primary field \cite{SZ}. The same comment as in the spin $1/2$ case applies again: there is no meaningful (unitary) dual description for the gravitino either.

\acknowledgments{ We thank C. Berger, P. de Smet and M. Ro\v{c}ek for conversations. We are especially grateful to P. Grassi and C. Zoubos for valuable discussions.
The present work was supported by the Research Foundation under NSF 
grant PHY-0098527.}

\end{document}